# Modelling the transmission and impact of Omicron variants of Covid-19 in different ethnicity groups in Aotearoa New Zealand

Samik Datta[1], Vincent X. Lomas[2,3], Nicole Satherley[4], Andrew Sporle[4], Michael J. Plank[2,5]

1. Earth Sciences New Zealand, Wellington, New Zealand.
2. School of Mathematics and Statistics, University of Canterbury, Christchurch, New Zealand.
3. Ngāi Tahu Research Centre, University of Canterbury, Christchurch, New Zealand.
4. iNZight Analytics Ltd, Auckland, New Zealand.
5. Te Pūnaha Matatini Centre of Research Excellence in Complex Systems, Auckland, New Zealand.

**Abstract**

Previous pandemics, including influenza pandemics and Covid-19, have disproportionately impacted Māori and Pacific populations in Aotearoa New Zealand. The reasons for this are multi-faceted, including differences in socioeconomic deprivation, housing conditions and household size, vaccination rates, access to healthcare, and prevalence of pre-existing health conditions. Many mathematical models that were used to inform the response to the Covid-19 pandemic did not explicitly include ethnicity or other socioeconomic variables. This limited their ability to predict, understand and mitigate inequitable impacts of the pandemic. Here, we extend a model that was developed during the Covid-19 pandemic to support the public health response by stratifying the population into four ethnicity groups: Māori, Pacific, Asian and European/other. We include three ethnicity-specific components in the model: vaccination rates, clinical severity parameters, and contact patterns. We compare model results to ethnicity-specific data on Covid-19 cases, hospital admissions and deaths between 1 January 2022 and 30 June 2023, under different model scenarios in which these ethnicity-specific components are present or absent. We find that differences in vaccination rates explain only part of the observed disparities in outcomes. While no model scenario is able to fully capture the heterogeneous temporal dynamics, our results suggest that differences between ethnicities in the per-infection risk of clinical severe disease is an important factor. Our work is an important step towards models that are better able to predict inequitable impacts of future pandemic and emerging disease threats, and investigate the ability of interventions to mitigate these.

## 1. Introduction

Aotearoa New Zealand used a combination of border and community control measures in 2020 and 2021 to minimise transmission of SARS-CoV-2 until high vaccine coverage could be achieved [1]. As a result, the country's rates of Covid-19 mortality and excess mortality were some of the lowest in the world [2]. However, despite the overall success of the response, the impact of Covid-19 was not evenly distributed and there were stark differences in health burden by ethnicity and socioeconomic deprivation [3,4]. In particular, Māori and Pacific populations had lower Covid-19 vaccination rates and experienced higher age-standardised rates of Covid-19 hospitalisation and Covid-19 mortality [5]. This mirrors international patterns where Indigenous populations, ethnic minorities, and socioeconomically deprived groups were often disproportionately affected [6,7,8].

Māori and Pacific people have higher average household size [9] and higher rates of employment in sectors where working from home is not possible [10]. These factors likely contribute to increased rates of exposure to respiratory diseases including Covid-19. Substantial differences in SARS-CoV-2 infection rates between ethnicities were also observed in England and in New York in 2020 [11,12]. In addition to differential rates of exposure, Māori and Pacific people also have higher prevalence of comorbidities [13,14] and poorer access to healthcare services [15–17], which likely means they are at higher risk of clinically severe disease [18,19]. However, disentangling the relative contributions of vaccination rates, clinical severity, and contact patterns to observed health inequities is challenging.

Mathematical models were used throughout the pandemic to inform Aotearoa New Zealand's response [20–23]. In most cases, these models did not model the interaction between disease dynamics and socioeconomic variables, such as ethnicity and deprivation. As a result, these models could not be used to predict differences in impact for different population groups, nor to answer policy questions about the effectiveness of interventions to mitigate inequities in the impact of Covid-19. Agent-based models offer one possible way to incorporate fine-scale information about population heterogeneity and its effect on disease transmission dynamics. Agent-based models were used during the pandemic in many countries including Aotearoa New Zealand [24]. However, although models of this type are highly valuable in some situations, they do have drawbacks. They typically have a large number of unknown parameter values which need to be estimated, they tend to be computationally intensive and challenging to calibrate to real epidemiological data in real time, and it can be difficult to identify which mechanistic assumptions are driving observed outcomes.

Limitations on the ability of models to quantify differential impacts of Covid-19 in important subpopulations were not unique to Aotearoa New Zealand. Policy-relevant models around the world frequently neglected socioeconomic heterogeneity in the population, either as a necessary model simplification or due to inadequate data or both. Improved quantification of the distributional impacts of disease has been identified as a key challenge for the next generation of infectious disease models [25,26]. There is an ethical obligation on policymakers and modellers to consider population heterogeneity and inequitable impacts in preparing for and responding to future pandemics [27,28]. Although understanding of the importance of factors commonly associated with inequitable disease impact is improving, our ability to disentangle the contributions of key drivers remains incomplete [29]. Mathematical models can help to provide insights in this regard, because they allow us to systematically test which hypotheses are consistent with observed data.

The aim of this study is to take an existing compartment-based model of Covid-19 that was used to inform the pandemic response in Aotearoa New Zealand [5] and investigate the effect of stratifying the population by ethnicity. The motivation for this is two-fold: (i) to develop models that are better equipped to support the public health response to future infectious disease threats and pandemics; and (ii) to improve understanding of the drivers of inequitable disease impacts, particularly the relative contributions of differences in contact rates, differences in vaccination rates, and differences in clinical severity (e.g. due to higher rates of comorbidity).

Subdividing the population into ethnicity groups requires estimates for relative contact rates between and within these groups. Directly measuring these contact rates is difficult and there is an absence of robust data. However, some progress can be made by comparing various simplified mechanisms for contact patterns, such as group-specific aggregate contact rates and varying levels of assortative mixing. We investigate scenarios in which vaccination rates, clinical severity parameters, and/or contact rates are ethnicity-specific. We compare these scenarios against a baseline model in which disease dynamics and impact are independent of ethnicity. By investigating the magnitude of the disparities across ethnicity groups, and comparing which scenarios fit best with observed ethnicity-specific data, we estimate the relative importance of these different factors in driving uneven disease impact.

## 2. Methods

### 2.1 Data

We obtained de-identified unit-record data from Te Whatu Ora (Health New Zealand) on reported cases of Covid-19 between 26 February 2020 and 31 December 2023. From this data, we calculated the daily number of reported Covid-19 cases, Covid-19 attributed hospital admissions and Covid-19 attributed deaths (see Supplementary Figure 1). These data were stratified into five-year age groups (with the last age group being over 75 years) and by Level 1 prioritised ethnicity in four groups: Māori, Pacific Peoples, Asian and European/other. We also obtained data on the daily number of first, second, third and fourth or subsequent Covid-19 vaccine doses stratified into the same age and ethnicity groups (see Supplementary Figure 2).

Reported Covid-19 cases were a mixture of healthcare-administered PCR tests and self-reported rapid antigen tests. Covid-19 hospitalisations were defined to be those where the patient was coded as receiving hospital treatment for Covid-19. Covid-19 deaths were defined to be deaths where Covid-19 was classified as the underlying cause of death or a contributory cause of death. We also obtained age-stratified data from Te Whatu Ora on the daily number of cases who filled a prescription for a Covid-19 antiviral treatment (either Paxlovid or molnupiravir, which were the two main treatments used in New Zealand).

For 8,505 of 2,547,479 reported cases (0.3%), ethnicity data was not available and we imputed ethnicity as follows. For each case with missing ethnicity data, we looked for other cases that were reported within $\pm 7$ days and had the same values for the following variables: five-year age band; number of vaccine doses received prior to the case report date; whether or not the case resulted in hospitalisation or death. Using these criteria, we found at least one matching case for 8,493 of the 8,505 cases with missing data. For the remaining 12 cases, we relaxed the criteria to match only on report date within $\pm 7$ days and five-year age band. For each case with missing ethnicity data, we selected one of the matching cases at random to use for imputation.

For the main set of model runs, we used the 2022 Health Service User (HSU) population dataset to specify the population size in each combination of ethnicity and five-year age group. However, this population data is known to underestimate population size for Māori and Pacific people, particularly in younger age groups [30]. To investigate the sensitivity of model results to this, we ran a second set of model simulations using population projections produced by Statistics NZ (projected population as at 31 December 2021 using the 2018 Census population as the base) according to assumptions agreed to by Te Whatu Ora [31]. Both sets of population data are shown in Supplementary Table 1.

**2.2 Model**

We extended a previously published compartment-based, age-structured model for the transmission dynamics of SARS-CoV-2 in the New Zealand population and the resulting Covid-19 hospitalisations and deaths between 1 January 2022 and 30 June 2023 [5]. We do not describe this model in detail again here. In brief, the model includes: time-varying vaccination rates based on data for the number of doses administered in each age group; waning immunity and reinfection; age-specific infection-fatality and infection-hospitalisation ratios; time-varying age-specific contact rates representing the effects of public health policy and behavioural change; and the impact of major new Omicron subvariants. At a given point in time, the model categorised the susceptible population in each age group into one of 14 susceptibility classes depending on the type and timing of the most recent immunising event (either previous infection or first, second, third or fourth or subsequent vaccine dose). Each susceptibility class had differing levels of immunity to infection and to severe disease or death. Waning immunity was modelled by transition to a susceptibility class with lower levels of immunity. Sixteen of the model parameters were previously estimated by fitting the model to data on reported daily cases, Covid-19-attributed hospital admissions and Covid-19-attributed deaths using approximate Bayesian computation [5]. These parameters control the relative overall contact rates over time as public health interventions were relaxed over the course of 2022, as well as the probabilities of detection, hospitalisation and death, the effectiveness of antiviral treatments, and the rate of immune waning (see Supplementary Table 2 for parameter values).

We extended this model by subdividing each compartment (representing a combination of age group and epidemiological status) into four groups representing people of Māori, Pacific, Asian and European/other ethnicity. We introduced the following ethnicity-specific effects into the model:

1. Ethnicity-specific vaccination rates using data for the daily number of first, second, third, and fourth or subsequent doses administered in each age and ethnicity group.

2. Ethnicity-specific clinical severity parameters (infection-hospitalisation and infection-fatality ratios).
3. Ethnicity-specific contact rates, quantified by an extended contact matrix $C$, where $C_{(i,d)(j,e)}$ is the average daily number of contacts that someone in age group $i$ and ethnicity $d$ has with people in age group $j$ and ethnicity $e$.

Note that while effects 1 and 3 above affect the transmission dynamics of the model by modifying the time-varying incidence of infections, effect 2 only affects the proportion of infections that go on to cause hospitalisation or death and not the underlying dynamics. We assumed that other model parameters such as vaccine effectiveness, rate of waning immunity, and case ascertainment rate were independent of ethnicity.

Our aim was to investigate the relative contributions of the effects above by comparing model outputs to ethnicity-stratified data, while accounting for background parameter uncertainty arising from the fit of the previous ethnicity-aggregated model to data [5]. We therefore treated the ethnicity-specific parameters as fixed in any given model scenario, and compared a range of hypothetical scenarios in which these effects were present or absent (see Table 1). This allows the mechanistic impact of different sources of inter-ethnicity variation to be directly assessed. An alternative approach would be to simultaneously account for uncertainty in the population-level and ethnicity-specific parameters. However, this would obscure the effect of parameters of interest so we do not attempt this here.

In the baseline scenario (Scenario 1), no ethnicity-specific model effects were included and the per capita vaccination rates in each age group were the same for all ethnicities. The only difference between the groups in Scenario 1 was the age structure of their respective populations. This scenario provided a benchmark against which to compare scenarios with different ethnicity-specific effects included.

Scenario 2 included ethnicity-specific vaccination data for the daily number of people receiving their first, second, third, or fourth or subsequent vaccine dose in each five-year age group and each ethnicity.

Scenarios 3-5 included additional effects on top of ethnicity-specific vaccination data. In scenarios 1-2, the infection hospitalisation ratio and infection fatality ratio were functions of age and immunity, but independent of ethnicity. In scenario 3, we allowed these ratios to also depend on ethnicity by applied ethnicity-specific odds ratios (see Table 1). These odds ratios were estimated by fitting logistic regression models to the data on Covid-19 hospitalisations and Covid-19 deaths, including age and vaccination status as predictor variables (see Supplementary Material sec. 1.1 for details). We applied the odds ratios shown in Table 1 to the infection-fatality rate and infection-hospitalisation rate in each age group, such that the average rate across ethnicities in that age group was the same as in the model without ethnicity stratification. Note that hospitalisation and death are modelled as independent processes and we do not make any assumption about what fraction of deaths occur in hospitalised patients.

Scenario 4 assumed that the clinical severity parameters were independent of ethnicity, and included an ethnicity-dependent mixing model. To implement this, we defined the extended contact matrix $C$ to represent assortative mixing (i.e. people are more likely to mix with others in the same ethnicity group) with ethnicity-specific relative contact rates (see Supplementary Material sec. 1.2 for details). Selecting parameter values for the ethnicity-specific mixing model is challenging as there are no direct data, such as social contact survey data, from which these can be estimated. We based our parameter choices on the results of Lomas et al. [32], who estimated an approximate value for the assortativity constant of $\varepsilon = 0.2$ using a similar method to Ma et al. [25]. In a scenario where Māori and Pacific ethnicities were assumed to have a 33% lower case ascertainment rate than the European/other ethnicity, Lomas et al. [32] estimated relative contact rates during the first Omicron wave of approximately 2.05, 3.32 and 0.89 for Māori, Pacific and Asian ethnicities respectively, relative to European/other. These estimates came from a non-age-structured model. The age-dependent contact matrix in our model already introduces some differences in contact rates between ethnicities due to their different age structures (Māori and Pacific populations have younger age structures than the European/other group). Accounting for these differences, the estimates of [32] translate to relative contact rates of 1.8, 2.8 and 0.8 for Māori, Pacific and Asian ethnicities respectively (see Table 1). Extended contact matrices for scenarios 1-4 are shown in Supplementary Figure 3.

Finally, scenario 5 included a combination of the clinical severity and mixing effects. In this scenario, we assumed the effect sizes were weaker than those in scenarios 3 and 4 where these effects were considered in isolation. We modelled weaker differences in clinical severity by setting the ethnicity-specific odds ratios for hospitalisation and death to the arithmetic mean of 1.0 and the value of the odds ratio in scenario 3. We based the ethnicity-specific relative contact rates on the scenario of [32] in which all groups were assumed to have equal case ascertainment rates, leading to relative contact rates of 1.1, 1.5 and 0.8 for Māori, Pacific and Asian ethnicities respectively (see Table 1).

These scenarios were designed to provide qualitative insights into the magnitudes of effect sizes that would be needed to explain the disparities between ethnicity groups under different hypotheses, rather than to provide definitive parameter estimates. In all scenarios, we used the set of parameter combinations previously reported by Datta et al. [5] by fitting the model without ethnicity stratification to data. We ran the model for the 18 month time period from 1 January 2022 to 30 June 2023, which included four epidemic waves of varying size. We report model results corresponding to the mean of the estimated posterior distribution for the fitted parameters and credible intervals (CrI) corresponding to the 95% best-fitting parameter combinations (i.e. the 95% of samples from the estimated joint posterior over parameters that have the smallest error relative to the data). For time-varying results, we present the 95% curvewise CrI, which is a band containing the 95% best-fitting trajectories.

Data and Matlab code along with instructions on how to run the model are available at [https://github.com/michaelplanknz/ethnicity-disease-model-public](https://github.com/michaelplanknz/ethnicity-disease-model-public).

| Scenario | Vaccination rates | Clinical severity | Mixing model |
|---|---|---|---|
| 1 | Independent of ethnicity | Independent of ethnicity | Proportionate mixing |
| 2 | Ethnicity-specific data | Independent of ethnicity | Proportionate mixing |
| 3 | Ethnicity-specific data | Ethnicity-specific odds ratios:<br>M P A<br>Hosp. 1.7 2.3 1.0<br>Death 1.7 1.6 0.6 | Proportionate mixing |
| 4 | Ethnicity-specific data | Independent of ethnicity | Assortative mixing ($\varepsilon = 0.2$)<br>Relative contact rates:<br>M P A<br>1.8 2.8 0.8 |
| 5 | Ethnicity-specific data | Ethnicity-specific odds ratios:<br>M P A<br>Hosp. 1.35 1.65 1.0<br>Death 1.35 1.3 0.8 | Assortative mixing ($\varepsilon = 0.2$)<br>Relative contact rates:<br>M P A<br>1.1 1.5 0.8 |

**Table 1. Model scenarios investigated.** Scenario 1 was the baseline model in which all ethnicities are effectively identical apart from the age structure. Scenarios 2-5 included ethnicity-dependent vaccination rates according to data. Scenarios 3, 4 and 5 investigated alternative combinations of ethnicity-dependent clinical severity parameters and ethnicity-dependent mixing. The clinical severity column shows the values of ethnicity-specific odds ratios for hospitalisation and death relative to the European/other group. The mixing model column shows the values of the assortativity constant ($\varepsilon$) and ethnicity-specific contact rates relative to the European/other group. Abbreviations: M = Māori, P = Pacific, A = Asian.

## 3. Results

In the baseline model (Scenario 1), model compartments were subdivided by ethnicity but no ethnicity-specific effects were included (i.e. equal vaccination rates and clinical severity across ethnicity groups and proportionate mixing). This model provided a reasonably good fit to the time series for cases, hospital admissions and deaths aggregated over age and ethnicity (Figure 1). Ethnicity-aggregated results were almost identical for all scenarios considered, with the exception of scenario 4 in which the first and second waves were slightly smaller. Similar observations apply to age-stratified results (Supplementary Figure 4). This confirms that the scenarios considered allow ethnicity-specific effects to be explored, while maintaining approximately consistent ethnicity-aggregated results.

We now explore ethnicity-specific results and focus primarily on hospitalisations and deaths as these are relatively insensitive to variations in surveillance effort and test-seeking behaviour over the modelled time period. While we include results on cases for completeness, and as a useful benchmark, these need to be interpreted with caution as they may be affected by unknown differences in case ascertainment between ethnicity groups. Figure 2 shows model results for the cumulative number of infections, cases, hospital admissions and deaths per 100,000 people from 1 January 2022 to 30 June 2023, in each ethnicity group and each model scenario, alongside ethnicity-specific data. The baseline model (Scenario 1) had notable differences between ethnicities due to their different age structure. Although the baseline model matched the ethnicity-specific data for reported cases reasonably well, it underestimated the hospitalisation and death rates for Māori and Pacific people, and overestimated these for the Asian ethnicity group.

Scenario 2 (ethnicity-dependent vaccination) improved the match with ethnicity-specific hospitalisation and death rates but only slightly. This was likely because, although total Māori and Pacific vaccination rates were lower than in the European/other and Asian ethnicities, differences were less pronounced in older age groups, where the majority of severe cases occurred (see Supplementary Figure 3). Therefore, differences in vaccination rates alone appear insufficient to explain the observed disparities among ethnicities.

Scenario 3 (ethnicity-dependent vaccination and clinical severity) provides a good match with the ethnicity-specific data for all four groups. Scenario 4 (ethnicity-dependent vaccination and mixing) underestimates reported cases in European/other and Asian ethnicities. As cases were primarily self-reported, this could be explained by differences in case ascertainment between ethnicities, which are not reflected in the model. Scenario 4 captures some of the ethnicity-specific trends in hospitalisation and death, but does not match as closely as scenario 3 does and in particular underestimates the hospitalisation rate for Pacific people and (to a lesser extent) Māori. Finally, scenario 5 (ethnicity-dependent vaccination, severity and mixing) provides a similar quality match as scenario 3. Overall, scenarios 3 and 5 provide a substantially better fit than the other scenarios to the cumulative data on ethnicity-specific hospitalisations and deaths, as measured by the total absolute relative error (Table 2).

Similar overall trends are seen in age- and ethnicity-stratified results (Supplementary Figures 5-7), although the quality of fit to the data is more variable. For most age-ethnicity combinations, the data lies within the credible intervals for scenarios 3 and 5 in Supplementary Figures 5-7), with the exception of deaths in under 60s. However, the absolute number of deaths in under 60s was relatively low, which means these results are more subject to the effect of stochastic fluctuations. The model ignores any interaction effects between age and ethnicity, which could also affect the quality of the fit at the age x ethnicity level.

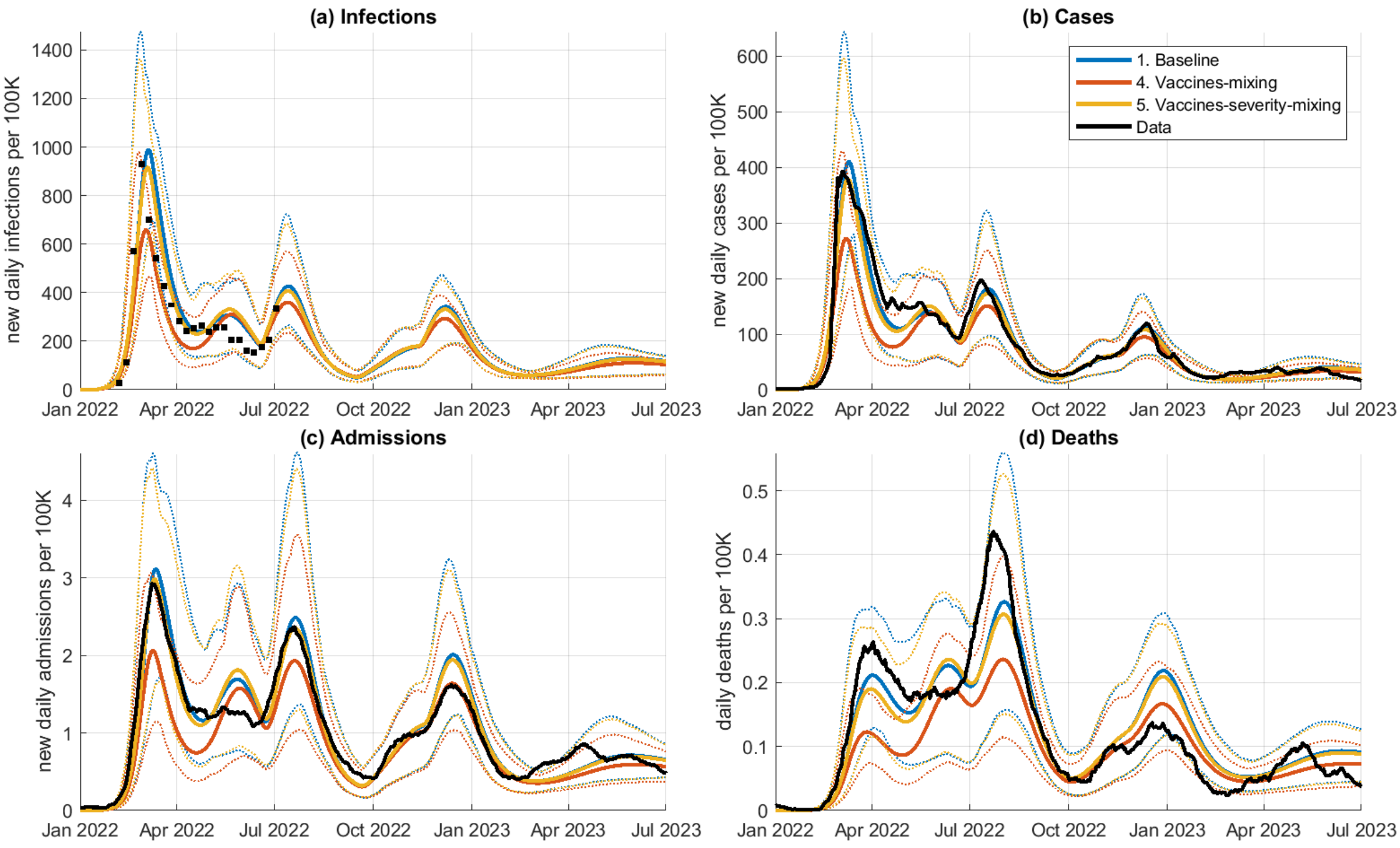

**Figure 1.** Key model outputs per 100,000 people aggregated over all ages and ethnicities for each model scenario, shown by different coloured lines (data shown by black points and lines): (a) new daily SARS-CoV-2 infections, (b) new daily reported Covid-19 cases, (c) new daily Covid-19 hospital admissions, (d) daily Covid-19 deaths. Solid curves show the model simulation under the posterior mean parameter values; dotted curves show the 95% CrI. To aid visual comparison, data for cases, admissions and deaths are shown as a moving average over a 7, 14 and 21-day window respectively. Note scenarios 2 and 3 are not shown because, when aggregated over ethnicities, they are identical to scenario 1.

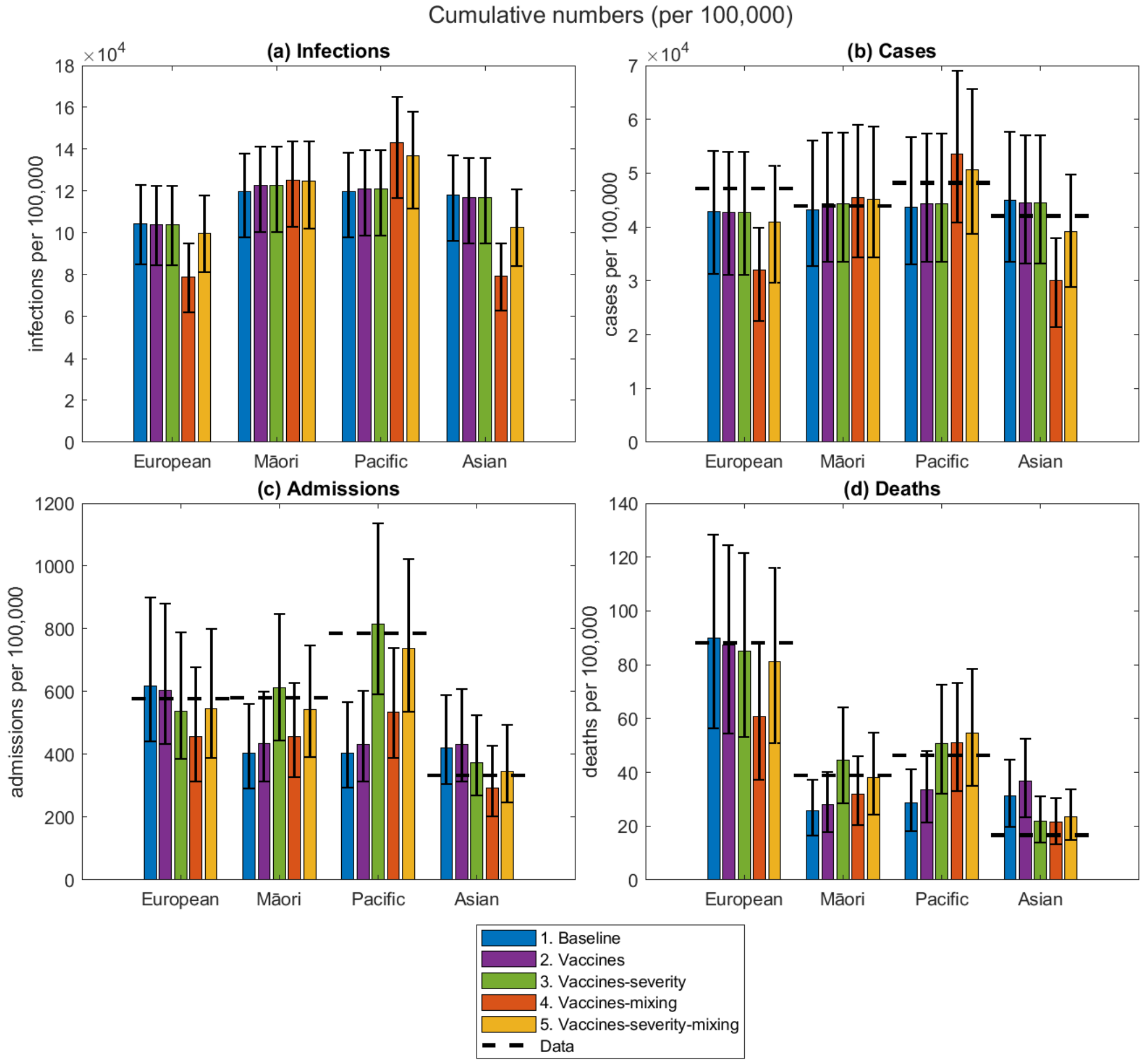

**Figure 2.** Cumulative number of SARS-CoV-2 infections, reported Covid-19 cases, Covid-19 hospital admissions and Covid-19 deaths per 100,000 people for each ethnicity, aggregated over all ages, from 1 January 2022 to 30 June 2023, comparing model scenarios (coloured bars) and data (horizontal red dashed lines). Coloured bars show the median; error bars show the 95% CrI. Note that no data are available for cumulative infections over this time period.

| Scenario | Relative error (admissions) | Relative error (deaths) |
|---|---|---|
| 1. Baseline | 1.117 | 1.608 |
| 2. Vaccines | 1.040 | 1.771 |
| 3. Vaccines-severity | 0.277 | **0.589** |
| 4. Vaccines-mixing | 0.865 | 0.886 |
| 5. Vaccines-severity-mixing | **0.225** | 0.697 |

**Table 2.** Total absolute relative error between ethnicity-specific model results and data for the cumulative numbers of hospital admissions and deaths, aggregated over all ages, from 1 January 2022 to 30 June 2023. Bold entries show the scenario with the lowest error. For each scenario, error was calculated as $\sum_e \left| x_e^{(\mathrm{model})} - x_e^{(\mathrm{data})} \right| / x_e^{(\mathrm{data})}$, where $x_e^{(\mathrm{model})}$ and $x_e^{(\mathrm{data})}$ are the cumulative number of outcomes in ethnicity group $e$ according to the median model estimate and the data respectively.

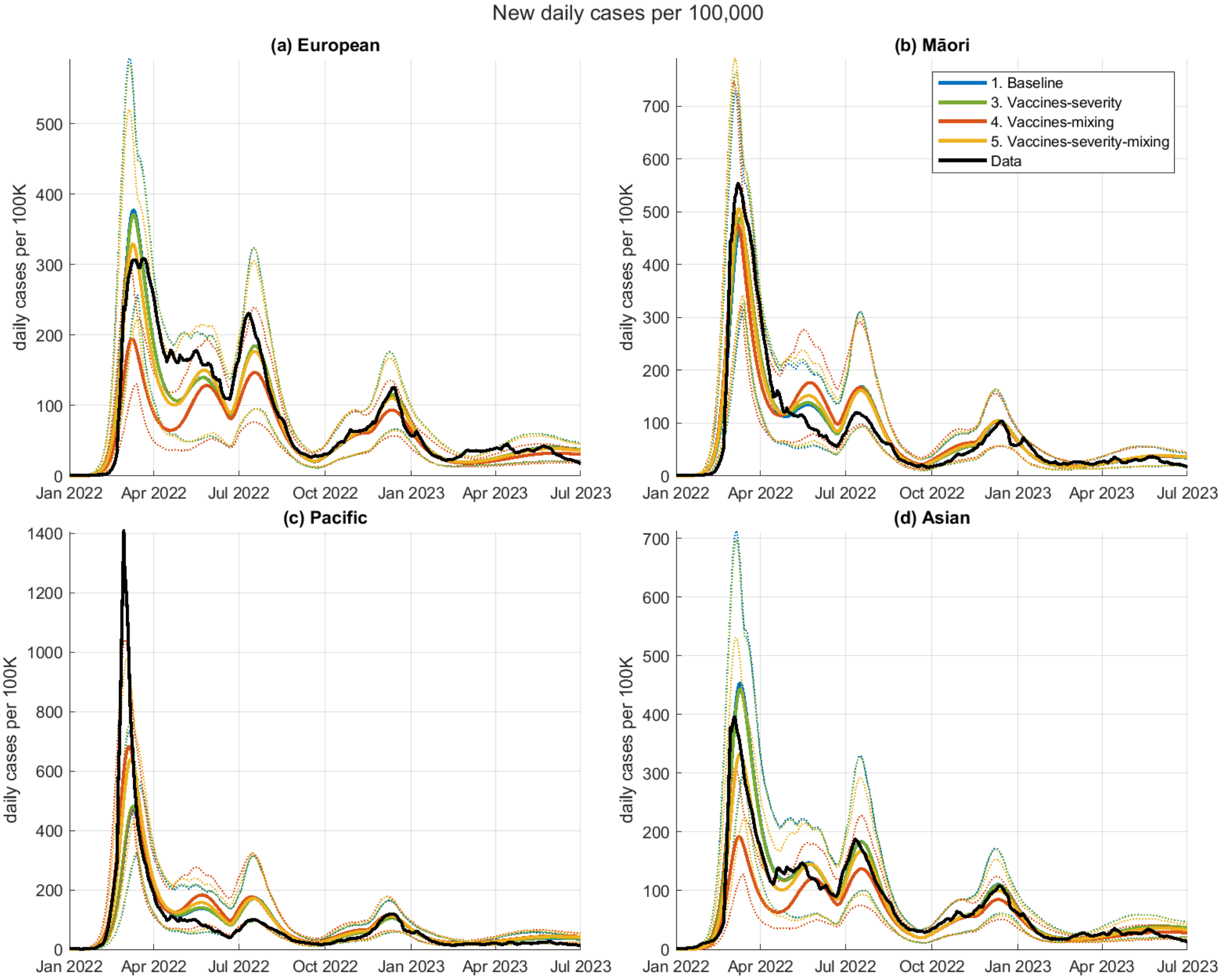

**Figure 3.** New daily reported Covid-19 cases per 100,000 people for each ethnicity, aggregated over all ages, comparing model scenarios 1, 3, 4 and 5 (different colours) to data (black lines). Scenario 2 is not shown as it is very similar to scenario 1. Solid curves show the model simulation under the posterior mean parameter values; dotted curves show the 95% CrI. To aid visual comparison, data for cases, admissions and deaths are shown as a moving average over a 7, 14 and 21-day window respectively.

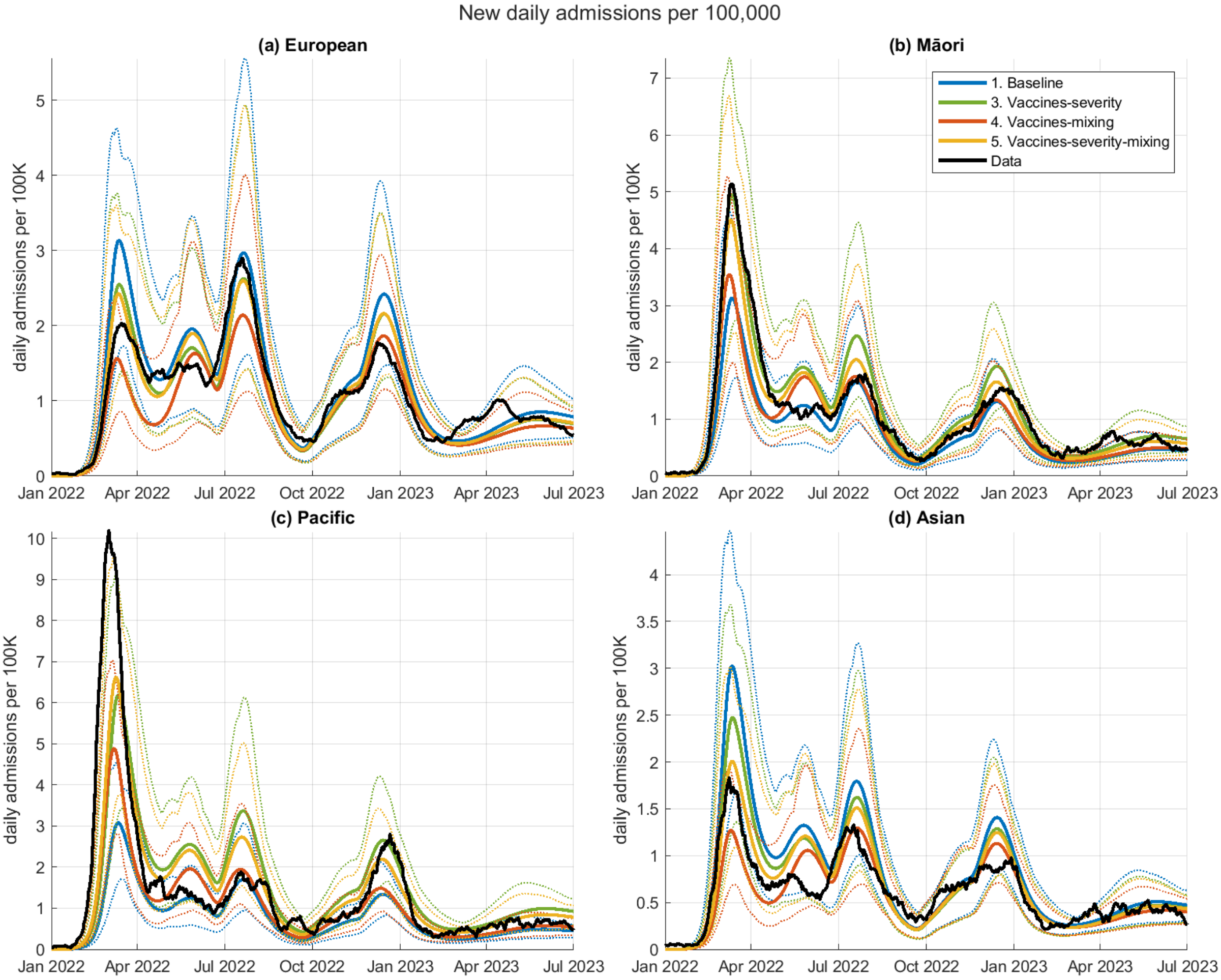

**Figure 4.** New daily Covid-19 hospital admissions per 100,000 people for each ethnicity, aggregated over all ages, comparing scenarios 1, 3, 4 and 5 (different colours) to data (black lines). Scenario 2 is not shown as it is very similar to scenario 1. Solid curves show the model simulation under the posterior mean parameter values; dotted curves show the 95% CrI. To aid visual comparison, data for cases, admissions and deaths are shown as a moving average over a 7, 14 and 21-day window respectively.

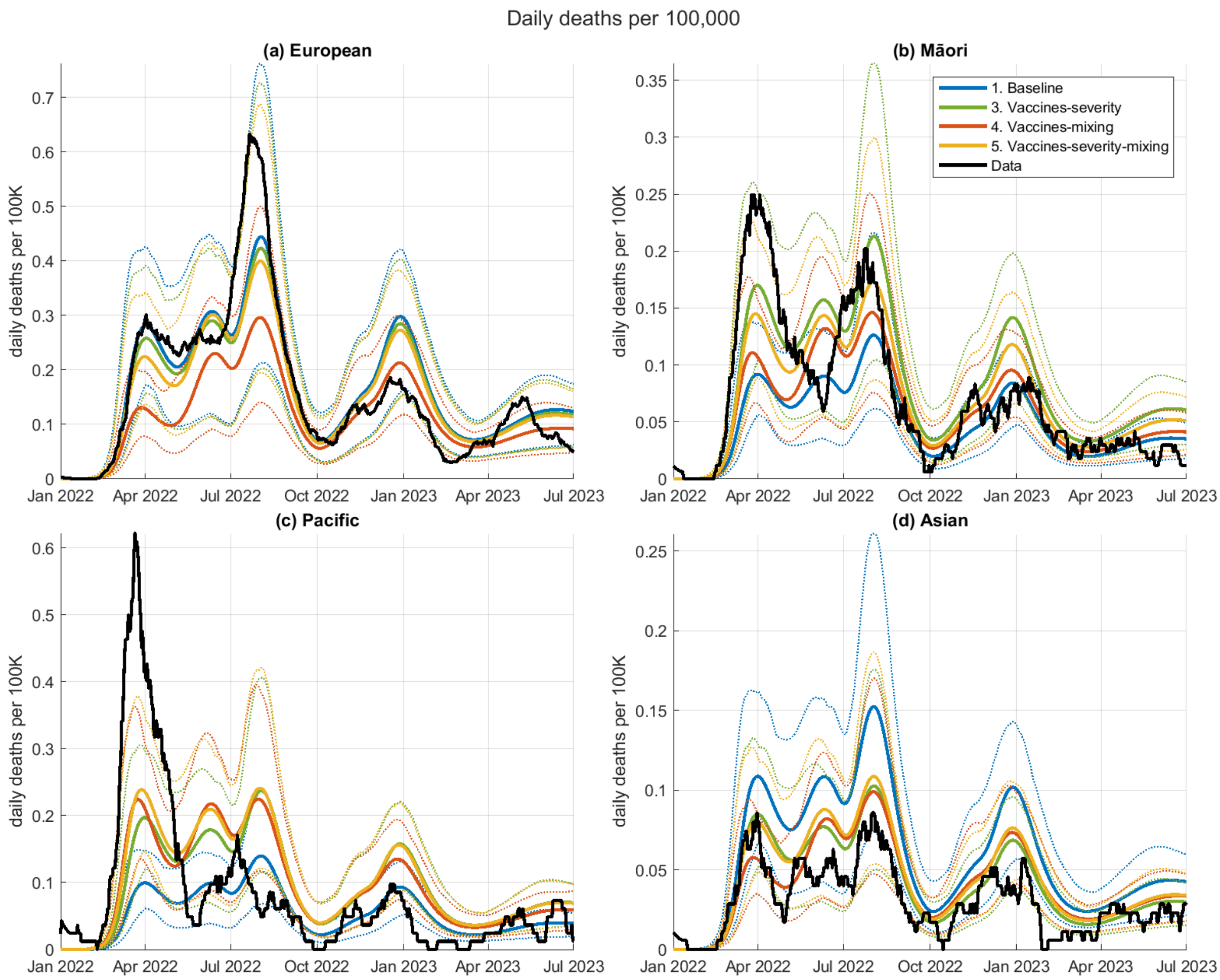

**Figure 5.** Daily Covid-19 deaths per 100,000 people for each ethnicity, aggregated over all ages, comparing scenarios 1, 3, 4 and 5 (different colours) to data (black lines). Scenario 2 is not shown as it is very similar to scenario 1. Solid curves show the model simulation under the posterior mean parameter values; dotted curves show the 95% CrI. To aid visual comparison, data for cases, admissions and deaths are shown as a moving average over a 7, 14 and 21-day window respectively.

Looking at time-dependent data shows that the first Omicron wave, which occurred in February-April 2022, disproportionately affected Māori and Pacific people, with a higher peak in reported cases (Figure 3), hospital admissions (Figure 4) and deaths (Figure 5). The second wave in July 2022 had a greater impact on people of European/other ethnicity, largely due to a shift in the age composition of cases towards older groups [20] (see Supplementary Material for age-stratified results).

As with the cumulative results in Figure 2, scenarios 1 (baseline model) fails to capture the disparities among ethnicity groups in time-dependent rates. In particular, it greatly underestimates the hospital admission and death rates in Māori and Pacific people, while overestimating these for Asian and European/other ethnicity groups (Figures 4-5).

None of the scenarios fully capture the differences between ethnicities in the dynamics of the first wave, in particular the very high peak seen in the Pacific population. Overall, consistent with the cumulative results, the best match is provided by scenarios 3 and 5. While the fit for these scenarios is far from perfect, it does capture the key ethnicity-specific trends while retaining a good fit to the aggregated data (see Figure 1). The time series for hospital admissions and deaths in each ethnicity group lie within the model's 95% CrI most of the time (with the exception of the first peak in the Pacific group as noted above). Scenario 4 provides a similar quality match to hospital admissions (Figure 4) but a poorer match to deaths (Figure 5).

Using the Statistics NZ population projections instead of the HSU dataset to determine population size by age and ethnicity gave qualitatively similar results (Supplementary Figure 7). The biggest difference was that, with the Statistics NZ projections, ethnicity-dependent vaccination alone (scenario 2) was sufficient to largely explain observed Māori hospitalisation and death rates, while scenarios 3 and 5 tend to overestimate these. In contrast, all of the scenarios underestimate Pacific hospitalisation rates.

## 4. Discussion

We have developed a compartment-based model of Covid-19 stratified by age and ethnicity. The model has three ethnicity-specific components: vaccination rates, clinical severity parameters (i.e. risk of severe disease or death following infection), and contact patterns. Ethnicity-specific vaccination rates were taken from vaccination records; ethnicity-specific clinical severity parameters were estimated independently of the transmission model using logistic regression models for the proportion of reported cases resulting in hospitalisation or death. We compared model outputs to ethnicity-stratified data on Covid-19 cases, hospital admissions and deaths in Aotearoa New Zealand between 1 January 2022 and 30 June 2023, under a range of counterfactual scenarios where each of the ethnicity-specific effects was present or absent. This allowed us to explore the relative contributions of different causal mechanisms to observed epidemiological disparities. These mechanisms are mediated via a complex and interacting network of socioeconomic, geographic and cultural factors, which did not attempt to explicitly model.

Our results showed that disparities in vaccination rates likely accounted for a relatively small part of the observed differences in Covid-19 impacts between ethnicities. This is consistent with the findings of Lomas et al. [32] who used a simpler model of the first Omicron wave in New Zealand. This finding is explained by the fact that these disparities in vaccination rates were relatively small in older age groups where most severe cases were concentrated. One caveat to this finding is that, when using population projections from Statistics NZ [31] instead of the HSU for population size data, disparities in vaccination rates had a bigger impact in the model on hospitalisation and death rates for Māori (but not Pacific people). The documented undercount of populations by the HSU is larger for Māori than for Pacific peoples [30], which may contribute to the larger difference in results by data source for Māori. There is a broader limitation in the availability of high quality and detailed ethnic population data in New Zealand in general outside of census years. This essentially limits yearly ethnic population data source options to the HSU and population projections. Administrative Population Census data published by Statistics NZ is a notable additional data source, but this data is currently available to 2022 only.

It is difficult to definitively disentangle differences in contact patterns and differences in clinical severity per infection with the data available. This is because, in the absence of reliable data on infection incidence (as opposed to self-reported cases), observed data could be explained by higher infection incidence resulting from higher contact rates in Māori and Pacific people, or by a higher per-infection risk of clinically severe outcomes. In mathematical terms, ethnicity-specific clinical severity parameters and ethnicity-specific contact rate parameters are not uniquely identifiable with the available data. Independent estimates of the relative infection rates of different ethnicities (e.g. estimated from an infection prevalence survey or seroprevalence data) would be needed to fully resolve this ambiguity.

If case ascertainment rates were similar across ethnicities, the most parsimonious explanation for the observed trends is that they were driven primarily by differences in clinical severity between ethnicities. It is difficult for the model to fully explain observed trends as a consequence of differences in mixing patterns alone. A modelling study [25] and social contact surveys [33,34] conducted in the US during the Covid-19 pandemic estimated that average contact rates typically differed between ethnicities by a factor of up to 2. We found that using comparable differences between ethnicities in average contact rates in the model was insufficient to explain observed disparities in hospitalisation and deaths. However, this does not preclude the possibility that the observed data could be due to a combination of lower case ascertainment, higher clinical severity, and higher contact rates in Māori and Pacific people relative to people of European ethnicity. In other words, observed disparities could have been due primarily to differences in per-infection risk, or to a combination of differences in per-infection risk and differences in contact rates, but are less likely to be due to differences in contact rates alone.

These findings are consistent with empirical studies of respiratory disease in New Zealand. Studies of severe disease caused by respiratory pathogens commonly show higher rates among Māori and Pacific populations [3,35]. Among patients hospitalised for Covid-19, Māori and Pacific patients had higher risk of complications including acute kidney injury and cardiac arrhythmia [18]. Data from the WellKiwis study shows that, in a highly tested cohort, differences in the incidence rate of acute respiratory infection (including mildly symptomatic cases) between ethnicity groups are relatively small [36], suggesting that differences in risk of severe illness are a bigger factor than differences in incidence of infection.

Our study has several important limitations. The ethnicity-specific model components investigated here are specific to the time period considered (1 January 2022 to 30 June 2023). During this time, contact patterns were still substantially affected by public health interventions and behavioural change associated with the Covid-19 pandemic. For example, interventions such as mask mandates, gathering size limits, testing recommendations, case and contact isolation requirements and work-from-home recommendations were still in place for the first wave and were progressively relaxed throughout 2022. It is probable that these interventions differentially affected contact rates in different ethnicities due to cultural and socioeconomic factors. For example, Māori and Pacific people have higher rates of employment in hospitality and the primary sector [10], where working-from-home is not possible. As a result, ethnicity-specific contact patterns cannot be translated to non-pandemic periods. Although it is likely that some of the same factors will be present in a future pandemic, and ethnicity-specific contact patterns may be qualitatively similar, the magnitude of differences between contact rates in different ethnicities may be different.

Similarly, quantitative differences in clinical severity will be pathogen-dependent. However, as historical trends indicate [37], higher rates of comorbidities in Māori and Pacific populations and broader social determinants of health mean it is highly likely these groups will be at higher risk of clinically severe disease from future emerging or re-emerging pathogens.

It is common for people to hold and report multiple ethnic identities in New Zealand which can create challenges and limitations for coding when mutually exclusive categories are required [38]. We used a prioritisation method which prioritised the coding in order of Māori ethnicity, followed by Pacific, Asian, and European/other. Around 10% of the Pacific population are also of Māori ethnicity, with multiple ethnic identification being much more common in younger age groups, meaning some of the Pacific population would have been coded under Māori ethnicity. As such, the data may not capture the entirety of the Pacific population, with an age-related bias that may undercount age groups with higher contact patterns due to work and education environments.

None of the model scenarios was able to capture the very high peak seen in the Pacific population in the first Omicron wave. This implies that other factors not included in the model played an important role, for example spatial heterogeneity. 62% of the national Pacific population lives in Auckland, New Zealand's largest city. Moreover, 34% live in South Auckland Local Board Areas [39], where the country's main international airport and the majority of its managed isolation and quarantine facilities were located. Rapid spread of SARS-CoV-2 in these highly connected communities early in the epidemic, which could include church-based communities, workplaces and extended family networks, could partially explain the discrepancy between model results and observed data for the Pacific population. In addition, our model made the simplifying assumption that the relative difference in contact rates between ethnicities were constant over time, which may not be the case. The first wave discrepancies noted above suggest that differences in contact rates likely played a major role early in the time period, but may have abated over time as pandemic interventions were relaxed and behavioural patterns returned to normal.

We have taken an exploratory, scenario-based approach to comparing model results with ethnicity-specific data on Covid-19 cases, hospital admission and deaths. It would be possible to take a more formal approach to parameter inference and model fitting, for example using a likelihood-based method [40] or approximate Bayesian computation [41]. However, given the issues outlined above in relation to parameter identifiability, we believe a more exploratory approach is appropriate, at this stage, to provide qualitative understanding of the relative contributions of different factors to disparities in observed data. This provides a starting point for further work to use more fine-grained models to provide a more nuanced understanding of the heterogeneous temporal dynamics.

Our work is an important step towards: (1) better understanding the drivers of inequities in the impact of pandemic respiratory pathogens, and (2) developing more nuanced models that can help prepare for and inform the public health response to future pandemics and emerging disease threats [27].

Remaining ambiguity as to the relative contribution of different drivers of observed inequities in outcomes from Covid-19 points to the need for data that is better able to resolve this. This could include representative ethnicity-specific data on infection incidence (as opposed to solely on more severe outcomes resulting in interaction with the healthcare system), for example from an infection prevalence survey or cohort study [42]. It could also include data on ethnicity-specific contact patterns from social contact surveys [33,34], ideally conducted during pandemic and non-pandemic periods. Accurately capturing ethnicity information about the respondent and their contacts would be difficult. However, having ethnicity information about respondents only would be sufficient to inform differences between ethnicities in overall contact rates. This would enable direct estimates of the parameters of our mixing model, rather than needing to infer these by fitting the model to noisy and potentially biased epidemiological data.

**Ethics statement**

This work did not require ethics approval.

**Data availability statement**

Data and Matlab code to reproduce the results in this article are available at https://github.com/michaelplanknz/ethnicity-disease-model-public, which contains full instructions for running the model. The ethnicity-specific epidemiological data cannot be shared for confidentiality reasons. This data can be requested from Te Whatu Ora (Health New Zealand) at: data-enquiries@tewhatuora.govt.nz.

**Funding**

The project "Improving models for pandemic preparedness and response: modelling differences in infectious disease dynamics and impact by ethnicity" (TN/P/24/UoC/MP) was funded by Te Niwha, the Infectious Diseases Research Platform – co-hosted by PHF Science and the University of Otago and provisioned by the Ministry of Business, Innovation and Employment, New Zealand. This research was supported by the Marsden Fund grant (24-UOC-020) managed by Royal Society Te Apārangi.

**Acknowledgements**

The authors are grateful to Mantaū Hauora (New Zealand Ministry of Health) and Te Whatu Ora (Health New Zealand) for supplying data in support of this work, in particular Laura Cleary, Fiona Wild and Elena Orsman. MJP would like to thank the Banff International Research Station for Mathematical Innovation and Discovery and the organizers and participants of the workshop Mathematical and Statistical Challenges in Post-Pandemic Epidemiology and Public Health (25w5369) held at Casa Matemática Oaxaca in Mexico. The authors are grateful to two anonymous peer reviewers for comments on an earlier version of this manuscript.

# Modelling the transmission and impact of Omicron variants of Covid-19 in different ethnicity groups in Aotearoa New Zealand

## Supplementary Material

## 1. Supplementary Methods

### 1.1 Ethnicity-specific clinical severity parameters

To estimate ethnicity-specific clinical severity parameters, we fitted logistic regression models to the data on Covid-19 hospital admissions and Covid-19 deaths for cases in adults reported between 15 January 2022 and 31 December 2022. Covid-19 testing during this period was predominantly via self-reported rapid antigen tests and so the results of these models will potentially be biased by differences in case ascertainment between ethnicity groups. We chose this period as it encompassed the three largest waves in the period of interest, but excluded later periods in which case ascertainment likely decreased further, which could confound the results. This period included 1,675,752 adult cases, of which 19,575 resulted in hospitalisation and 3391 resulted in death.

We specified the regression model as:

$$logit(p) = \beta_0 + \beta_1(age) + \beta_2(number\ of\ vaccine\ doses) + \beta_3(ethnicity), \qquad \text{(S1)}$$

where $p$ is the probability of the outcome of interest (either Covid-19 attributed hospitalisation or Covid-19 attributed death) following a positive Covid-19 test. Age was a continuous variable. Number of vaccine doses was a categorical variable, either 0, 1, 2, or 3 or more doses prior to the case report date. Ethnicity was a categorical variable, either Māori, Pacific, Asian or European/other. We chose this model specification to produce estimates of ethnicity-dependent severity that control for the effects of age and vaccination, because these variables are explicitly included in the main model. We did not control for other factors such as pre-existing health conditions because these variables partly mediate the relationship between ethnicity and risk of severe disease.

To incorporate ethnicity-dependent clinical severity into the dynamic model, we applied the estimated odds ratios for ethnicity (see Supplementary Table 1) to the infection-hospitalisation ratio (IHR) and infection-fatality ratio (IFR) parameters within each age group. Using $x_e$ to denote the IHR (or IFR) for ethnicity group $e$ in a given age group, we have

$$x_e = \frac{x_O OR_e}{x_O OR_e + 1 - x_O},$$

where $x_O$ is the value for the European/other group and $OR_e$ is the odds ratio for ethnicity $e$ relative to the European/other group. To maintain the same overall IHR (or IFR) $\underline{x}$ within each age group as in the model without ethnicity, we require that

$$\underline{x} = \sum_e p_e x_e ,$$

where $p_e$ is the proportion of that age group that is in ethnicity group $e$. These two equations enabled the ethnicity-specific IHR and IFR to be calculated in each age group (see Supplementary Table 2). Note that the values in Supplementary Table 2 show the IHR and IFR for individuals who are fully susceptible. These rates are reduced for people in compartments that have been vaccinated or previously infected according to the immunity model.

## 1.2 Ethnicity-specific contact rates

We modelled ethnicity-specific contact patterns via two simplified mechanisms: assortative mixing, quantified by a parameter $\varepsilon \in [0,1]$; and ethnicity-specific contact rates, specified by a vector of relative contact rates $c$. For simplicity we assumed that these mechanisms and parameters operated independently of age. This enabled us to define the extended contact matrix $C_{(i,d)(j,e)}$, representing the average daily number of contacts that someone in age group $i$ and ethnicity $d$ has with people in age group $j$ and ethnicity $e$, as:

$$C_{(i,d)(j,e)} = (1-\varepsilon)\frac{C_{ij} N_i N_{j,e} c_d c_e}{\sum_{d',e'} N_{i,d'} N_{j,e'} c_{d'} c_{e'}} + \varepsilon \frac{C_{ij} N_i N_{j,e} c_d \delta_{de}}{N_d \sum_{d'} \frac{N_{i,d'} N_{j,d'} c_{d'}}{N_{d'}}},$$

where $C_{ij}$ is the age-only contact matrix (i.e. the average daily number of contacts that someone in age group $i$ has with people with age group $j$, regardless of ethnicity), $N_{i,d}$ is the population size in age group $i$ and ethnicity $d$, $N_i = \sum_d N_{i,d}$ is the total population size in age group $i$, and $N_d = \sum_i N_{i,d}$ is the total population size in ethnicity group $d$.
The above definition of $C_{(i,d)(j,e)}$ preserves two important properties of the mixing model: (1) the average total number contacts that someone in age group $i$ has with people with age group $j$, regardless of ethnicity, is equal to the original, age-only contact matrix $C_{ij}$; (2) the total number of contacts between group $(i,d)$ and $(j,e)$, given by $N_{i,d} C_{(i,d)(j,e)}$, forms a symmetric matrix. The case $\varepsilon = 0$ corresponds to proportionate mixing, while $\varepsilon = 1$ corresponds to mixing strictly within ethnicity groups only, while values of $\varepsilon$ between 0 and 1 interpolate between these extremes. For our model, $C_{(i,d)(j,e)}$ was a $64 \times 64$ matrix, corresponding to the combination of 16 age groups with 4 ethnicity groups (see Supplementary Figure 1).

## 1.3 Transmission model

The transmission model consisted of a system of ordinary differential equations for the susceptible ($S$), exposed ($E$), infectious with symptoms ($I$), infectious without symptoms ($A$), and immune ($R$) compartments in each age ($i$), ethnicity ($d$) and susceptibility ($k$) class (see Datta et al. for details). The force of infection on susceptible class $S_{idk}$ was given by

$$\lambda_{idk}(t) = \frac{\alpha u_i r_k}{N_{id}} \sum_{j,e,l} C_{(j,e)(i,d)} (I_{jel} + \tau A_{jel}),$$

where $r_k$ is the relativity susceptibility of class $k$ (modelling the effects of partial immunity to infection due to vaccination or prior infection), $u_i$ is the relative susceptibility of age group $i$, $\tau$ is the relative infectiousness of asymptomatically infected individuals, and $\alpha$ is a constant representing the probability of transmission following contact between an infectious and a fully susceptible.

## 2. Supplementary Tables and Figures

| Age (yrs) | HSU data Euro | Māori | Pacific | Asian | **Total** | Statistics NZ data Euro | Māori | Pacific | Asian | **Total** |
|---|---|---|---|---|---|---|---|---|---|---|
| 0-4 | 142182 | 77999 | 31043 | 59436 | **310660** | 128050 | 86210 | 28460 | 56350 | **299070** |
| 5-9 | 154933 | 83904 | 33428 | 60679 | **332944** | 145780 | 86680 | 30590 | 58370 | **321420** |
| 10-14 | 180257 | 81845 | 35600 | 47777 | **345479** | 163700 | 93300 | 33690 | 48080 | **338770** |
| 15-19 | 172885 | 70634 | 33058 | 42417 | **318994** | 165020 | 80180 | 32100 | 45000 | **322300** |
| 20-24 | 180625 | 67274 | 36098 | 50796 | **334793** | 170330 | 73630 | 34100 | 53040 | **331100** |
| 25-29 | 202412 | 63189 | 35134 | 81931 | **382666** | 188810 | 68550 | 31670 | 79210 | **368240** |
| 30-34 | 213962 | 58429 | 31227 | 100897 | **404515** | 197760 | 62020 | 27760 | 96740 | **384280** |
| 35-39 | 194971 | 44994 | 26302 | 92630 | **358897** | 182530 | 49920 | 22600 | 89940 | **344990** |
| 40-44 | 189289 | 41818 | 24090 | 69003 | **324200** | 179530 | 46420 | 20790 | 67460 | **314200** |
| 45-49 | 210152 | 42850 | 22152 | 50227 | **325381** | 203350 | 47900 | 20180 | 49260 | **320690** |
| 50-54 | 230953 | 42786 | 21194 | 41711 | **336644** | 223400 | 47030 | 19640 | 41660 | **331730** |
| 55-59 | 230375 | 38895 | 18718 | 36311 | **324299** | 224070 | 42630 | 17350 | 37040 | **321090** |
| 60-64 | 224999 | 32840 | 14553 | 31064 | **303456** | 217950 | 35310 | 13080 | 31310 | **297650** |
| 65-69 | 195633 | 22930 | 10968 | 28542 | **258073** | 190850 | 25350 | 9770 | 26600 | **252570** |
| 70-74 | 178960 | 15459 | 7788 | 18604 | **220811** | 176990 | 16890 | 6780 | 17620 | **218280** |
| 75+ | 304679 | 16057 | 9362 | 21736 | **351834** | 299320 | 18650 | 8380 | 21500 | **347850** |
| **Total** | **3207267** | **801903** | **390715** | **833761** | **5233646** | **3057440** | **880670** | **356940** | **819180** | **5114230** |

**Supplementary Table 1.** Population size data stratified by age and ethnicity according to the HSU dataset (used in the main model runs) and the Statistics NZ data set (used in the sensitivity analysis).

| Parameter | Posterior median [95% CrI] |
|---|---|
| Outbreak start date | 17 [15, 21] days |
| $R_e(t)$ in period 1 | 2.24 [2.07, 2.44] |
| Period 1-2 ramp start date | 69 [63, 73] days |
| Period 1-2 ramp window | 56 [36, 74] days |
| $R_e(t)$ in period 2 | 3.69 [2.97, 4.24] |
| Period 2-3 ramp start date | 258 [252, 262] days |
| Period 2-3 ramp window | 11 [1, 19] days |
| Period 2-3 factor increase in $R_e(t)$ | 1.22 [1.11, 1.29] |
| Contact matrix relaxation factor | 0.49 [0.07, 0.78] |
| Contact matrix relaxation window | 69 [50, 89] days |
| Testing probability multiplier | 1.00 [0.81, 1.19] |
| IFR probability multiplier | 0.80 [0.52, 1.21] |
| IHR probability multiplier | 1.26 [0.92, 1.48] |
| Immunity waning rate | 0.0052 [0.0037, 0.0067] ($day^{-1}$) |
| BA.5 immune evasion | 0.37 [0.13, 0.63] |
| Antiviral effect on IFR | 0.50 [0.41, 0.60] |

**Supplementary Table 2.** Table of estimated posterior values (median and 95% credible interval) for the fitted parameters of the model without ethnicity stratification. Note model runs were performed using samples from the estimated joint posterior distribution over fitted parameters, while the table above only shows marginal statistics for each individual parameter. See Ref. [5] for details. Abbreviations: $R_e(t)$ – time-varying reproduciotn number in a fully susceptible population; IFR – infection fatality ratio. Note all 'start date' parameters are shown in days relative to 1 January 2022.

| Variable | Hospitalisation odds ratio | Death odds ratio |
|---|---|---|
| Age (years) | 1.07 [1.07, 1.07] | 1.15 [1.15, 1.15] |
| Ethnicity | | |
| European/other | (baseline) | |
| Māori | 1.73 [1.66, 1.81] | 1.68 [1.48, 1.91] |
| Pacific | 2.27 [2.17, 2.39] | 1.58 [1.36, 1.85] |
| Asian | 1.03 [0.98, 1.09] | 0.61 [0.50, 0.73] |
| Number of doses | | |
| 0 | (baseline) | |
| 1 | 1.28 [1.13, 1.46] | 1.19 [0.85, 1.67] |
| 2 | 0.46 [0.44, 0.49] | 0.57 [0.49, 0.67] |
| 3 or more | 0.24 [0.23, 0.25] | 0.25 [0.22, 0.28] |

**Supplementary Table 3.** Odds ratios (with 95% confidence intervals) for Covid-19 hospitalisation and Covid-19 death following a positive test for age, ethnicity (relative to the European/other group), and number of Covid-19 vaccine doses received prior to the case report date (relative to unvaccinated). Estimates are from logistic regression models for all Covid-19 cases among adults reported between 15 January 2022 and 31 December 2022 ($n$ = 1,675,752) - see Eq. (S1).

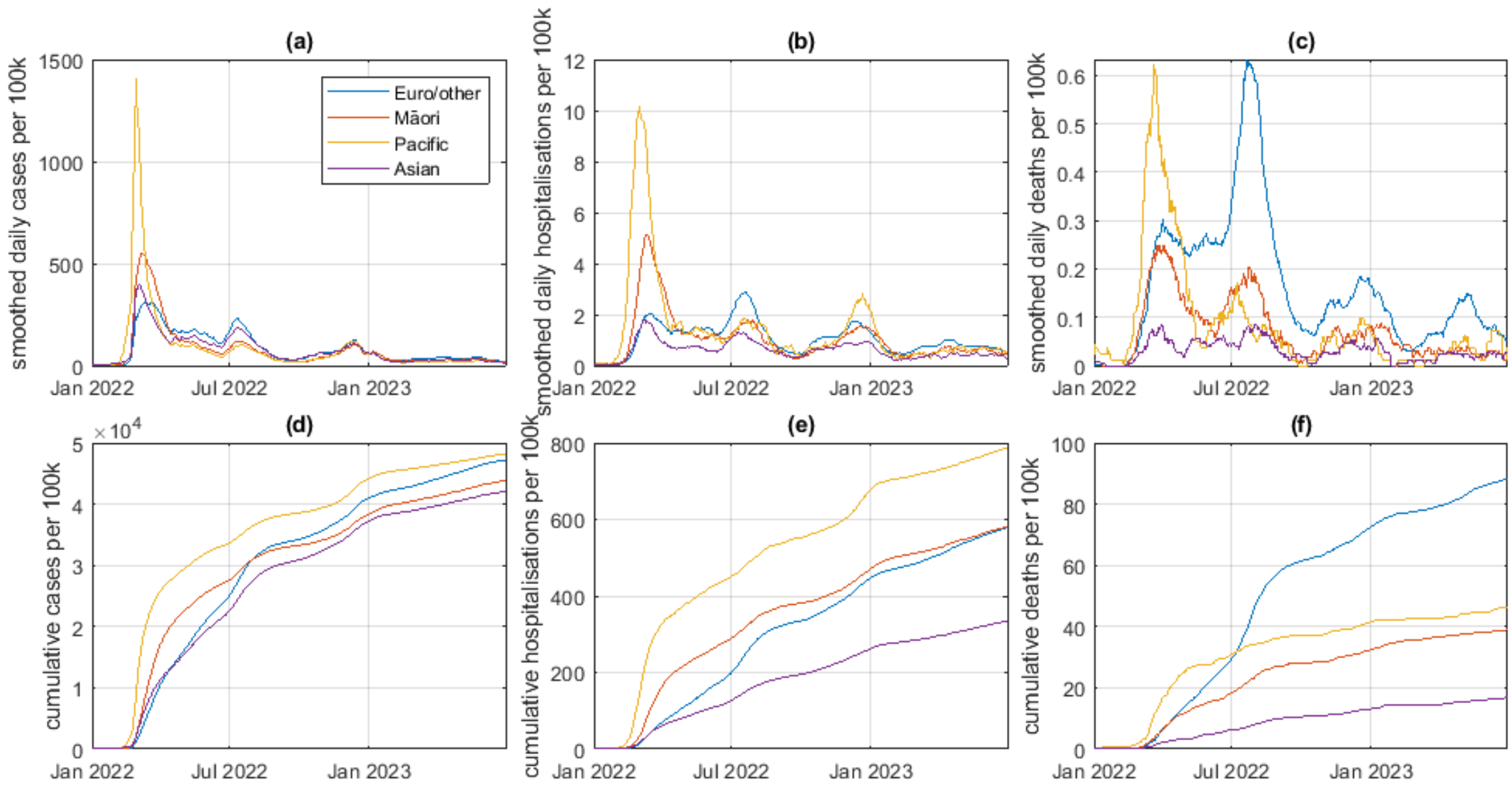

**Supplementary Figure 1.** Te Whatu Ora (Health New Zealand) data on the daily rates per 100,000 in each ethnicity group of: (a) reported Covid-19 cases (7-day moving average); (b) new Covid-19 hospital admissions (14-day moving average); (c) Covid-19 deaths (21-day moving average). Cumulative rates per 100,000 of: (d) reported Covid-19 cases; (e) Covid-19 hospital admissions; (f) Covid-19 deaths.

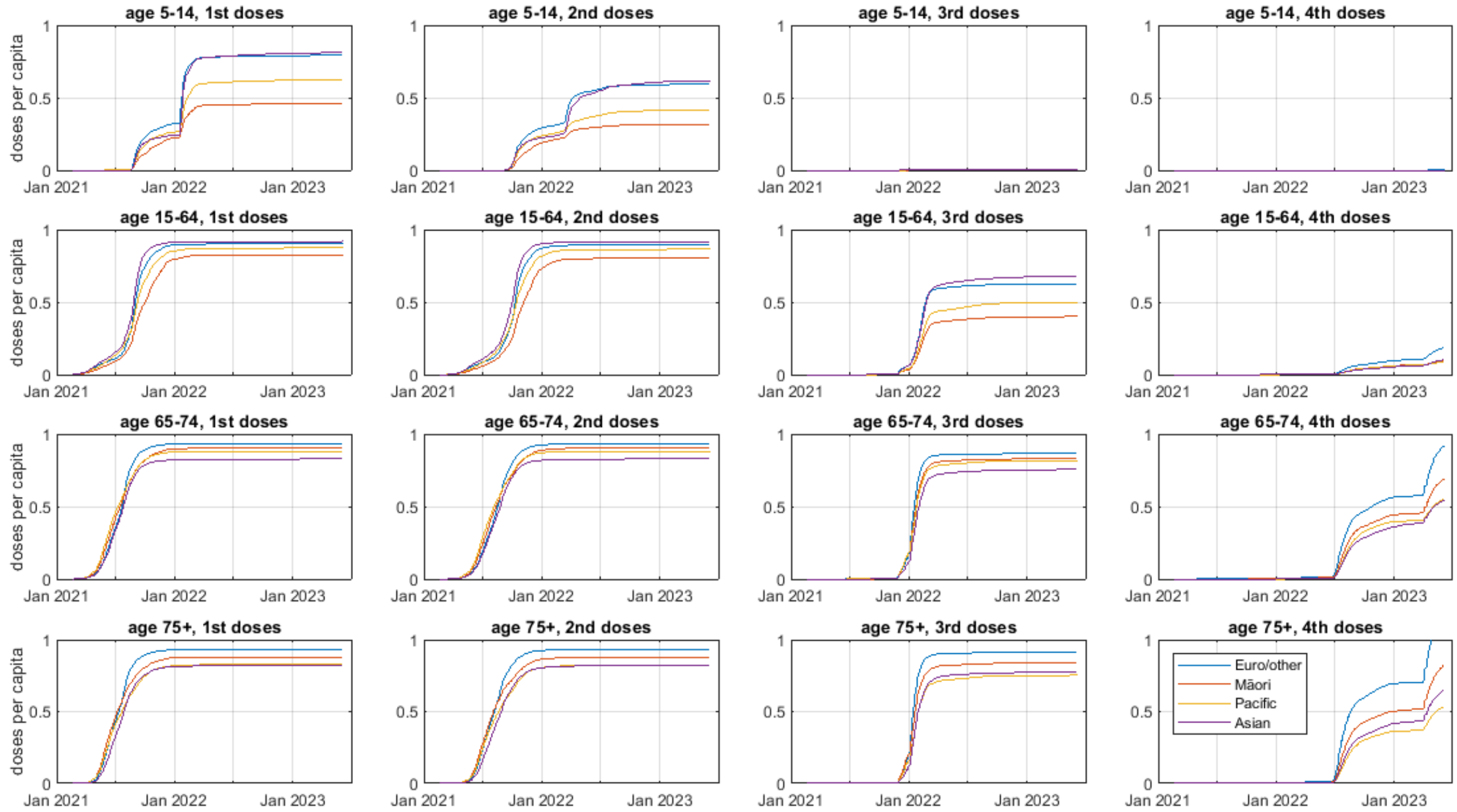

**Supplementary Figure 2.** Te Whatu Ora (Health New Zealand) data for the cumulative number of first, second, third and fourth or subsequent Covid-19 vaccine doses per capita in each ethnicity group and in age bands 5-14 years old, 15-64 years old, 65-74 years old and 75 years old and over. Note the per capita number of first doses, second doses and third doses is always between 0 and 1; the per capita number of fourth or subsequent doses can exceed 1, for example if more than 50% of the population group have received five doses.

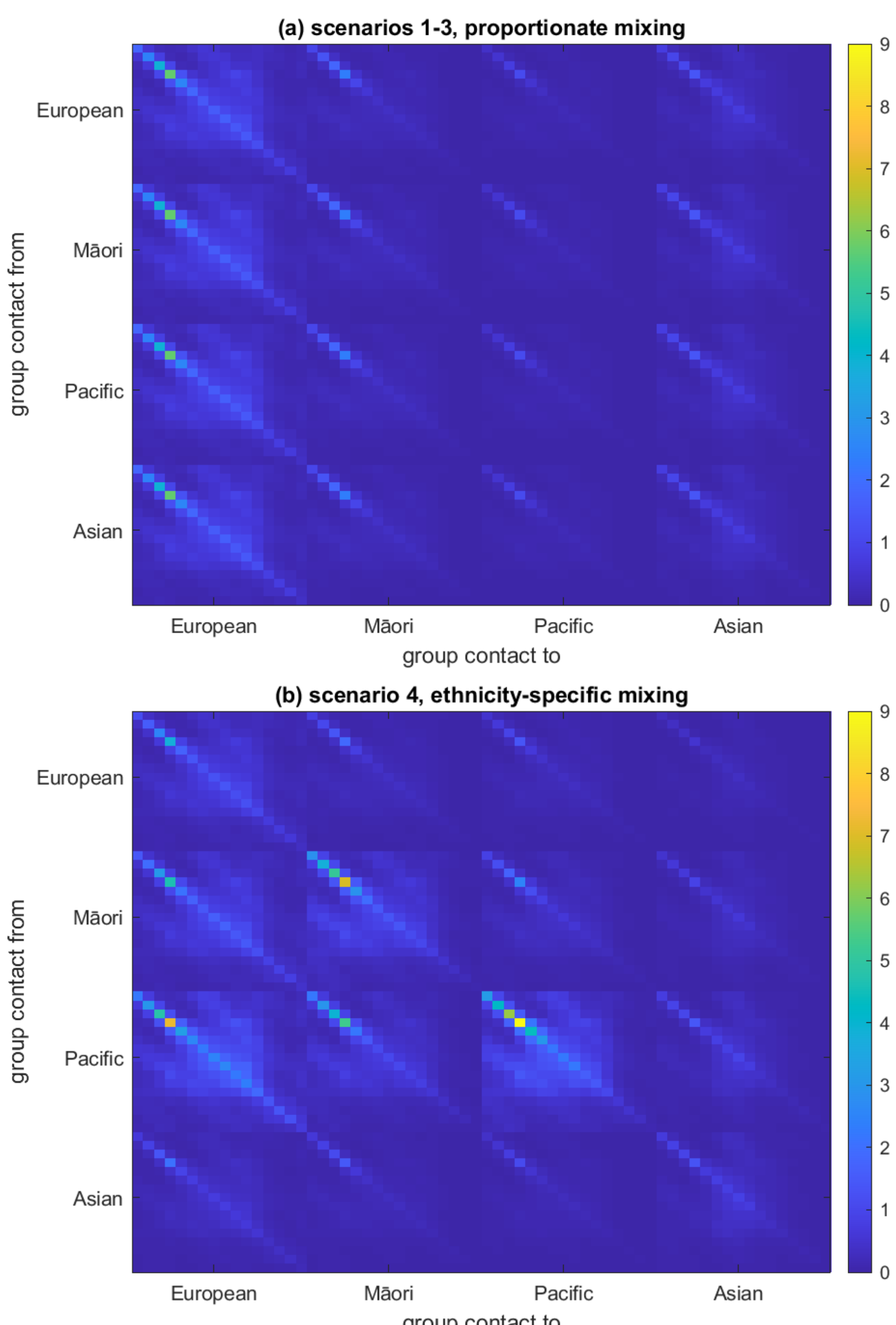

**Supplementary Figure 3.** Contact matrices used in the model for: (a) scenarios 1-3 (proportionate mixing among ethnicities); (b) scenario 4 (ethnicity-specific relative contact rates and assortative mixing among ethnicities – see Table 1). Each panel shows a 64 x 64 matrix, which consists of a 4 x 4 block, where each block consists of a 16 x 16 submatrix quantifying contact rates between the 16 age groups. Note that contacts to people of European/other ethnicity tend to occur more frequently as this is the largest ethnicity group (i.e. European block column tends to have higher values than the other block columns). In panel (b), note that: the Māori and Pacific groups have more contacts than in panel (a) (i.e. Māori and Pacific block rows have higher values in (b) than in (a)) because they have higher relative contact rates; and within-ethnicity contacts tends to be more frequent (diagonal blocks tend to have higher values in (b) than in (a)) due to assortative mixing.

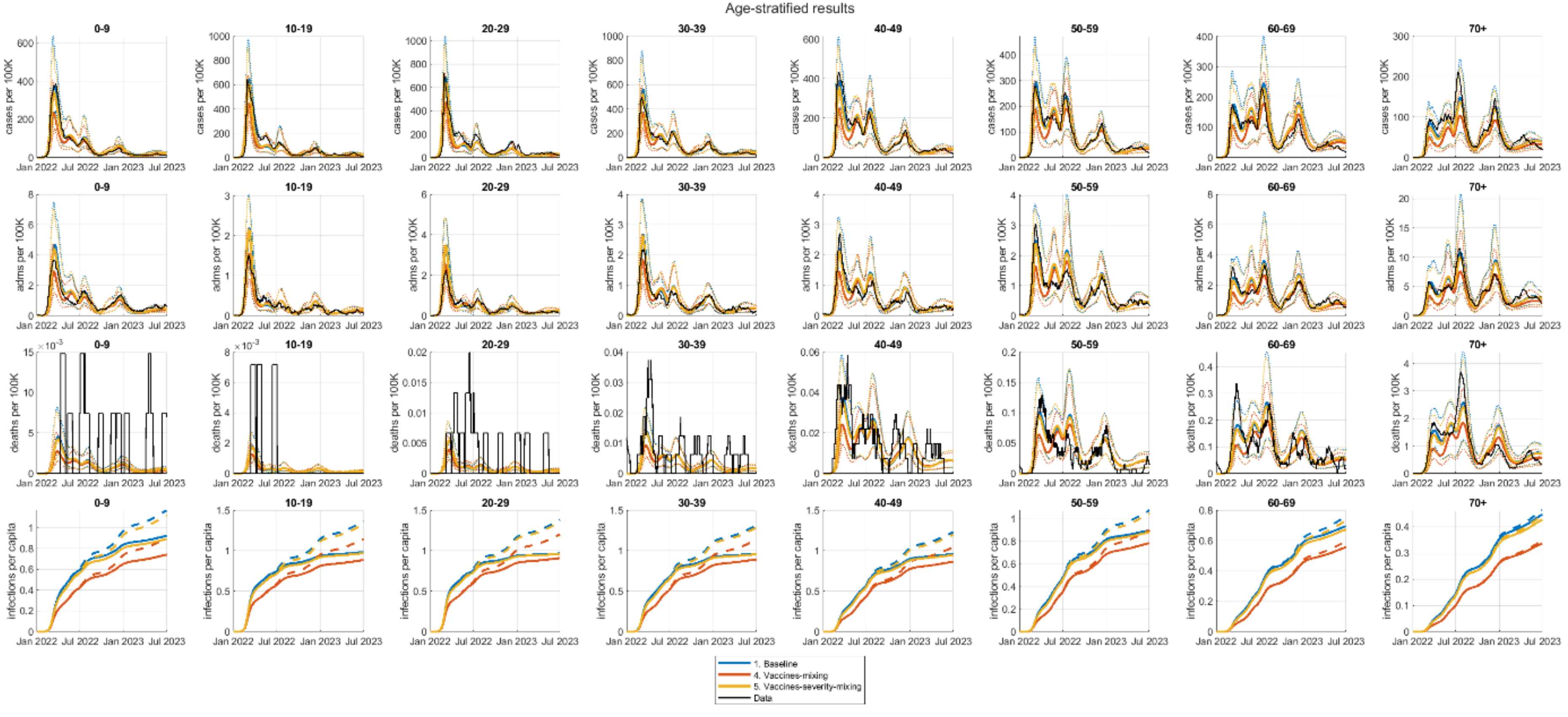

**Supplementary Figure 4.** Key model outputs per 100,000 people in ten-year age groups, aggregated over all ethnicities. Model scenarios are shown by different coloured lines; data shown by black lines. First row: new daily reported Covid-19 cases. Second row: new daily Covid-19 hospital admissions. Third row: daily Covid-19 deaths. Fourth row: cumulative infections per capita. Solid curves show the model simulation under the posterior mean parameter values; dotted curves show the 95% CrI. In the fourth row of plots (cumulative infections per capita), the solid curves show first infections and the dashed curves show all infections (first infections plus reinfections). To aid visual comparison, data for cases, admissions and deaths are shown as a moving average over a 7, 14 and 21-day window respectively. Note results for scenarios 2 and 3 are not shown because, when aggregated over ethnicities, they are identical to scenario 1.

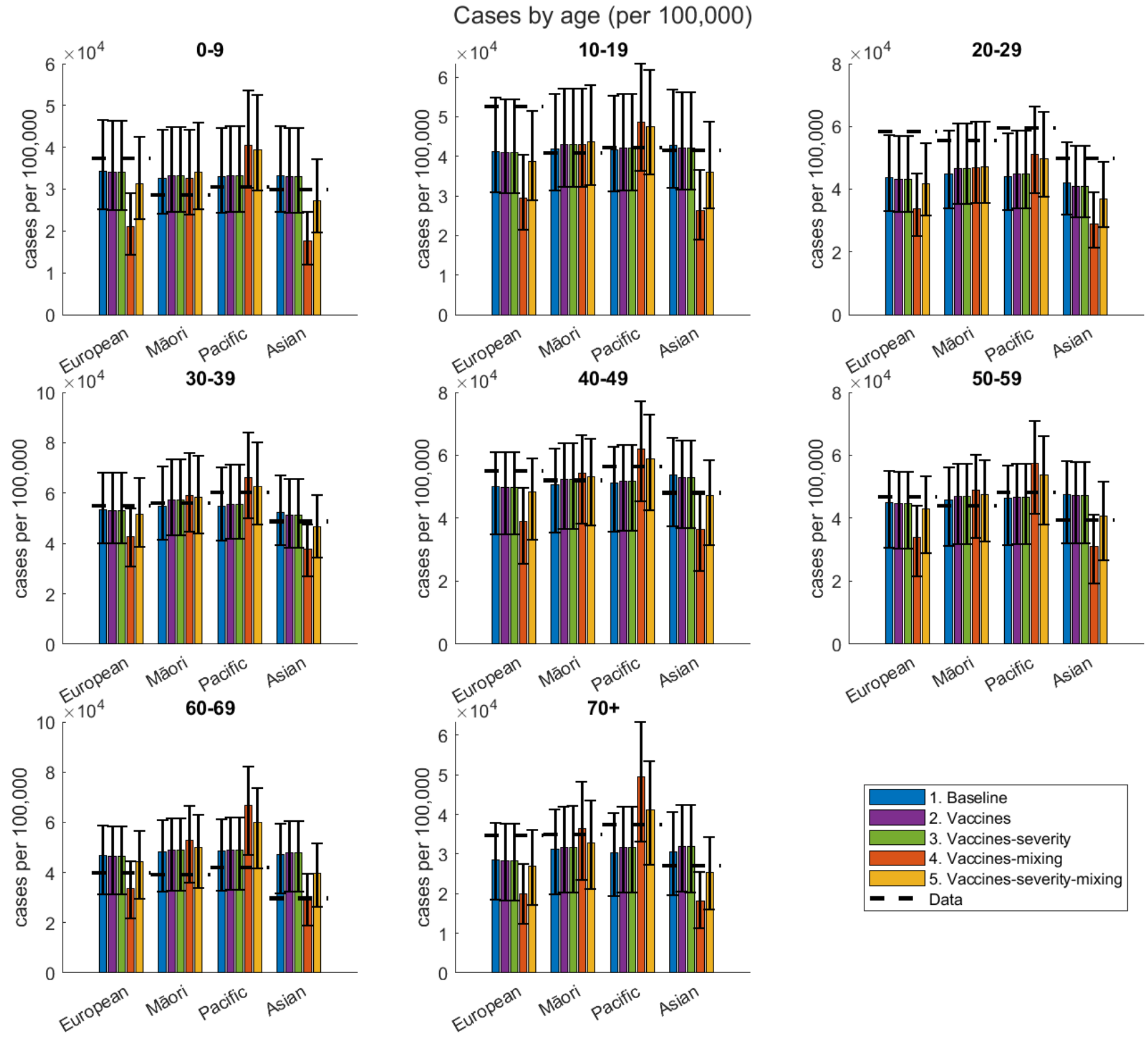

**Supplementary Figure 5.** Cumulative number of reported Covid-19 cases per 100,000 people in 10-year age groups from 1 January 2022 to 30 June 2023, comparing model scenarios (coloured bars) and data (horizontal red dashed lines). Coloured bars show the median; error bars show the 95% CrI.

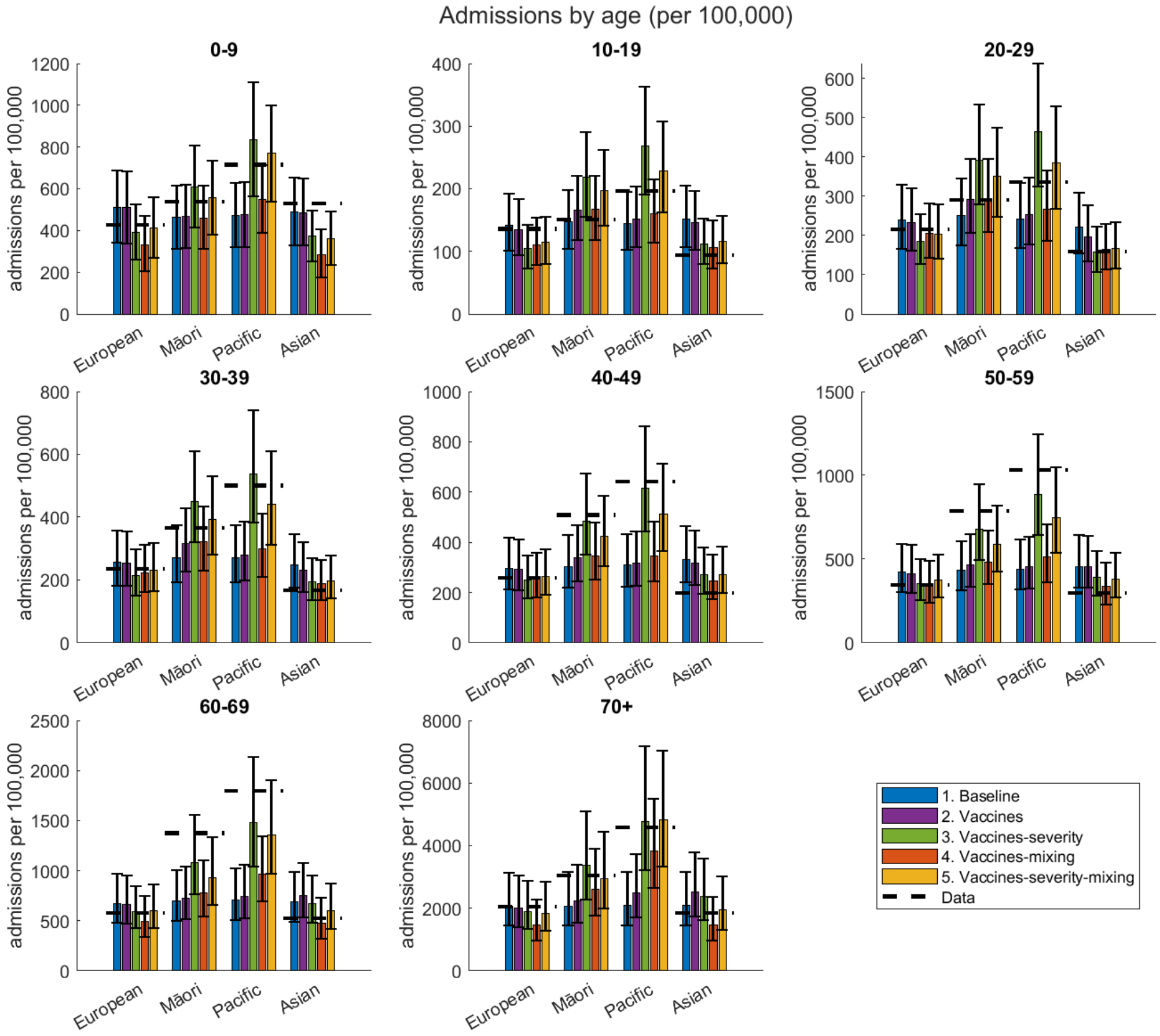

**Supplementary Figure 6.** Cumulative number of Covid-19 hospital admissions per 100,000 people in 10-year age groups from 1 January 2022 to 30 June 2023, comparing model scenarios (coloured bars) and data (horizontal red dashed lines). Coloured bars show the median; error bars show the 95% CrI.

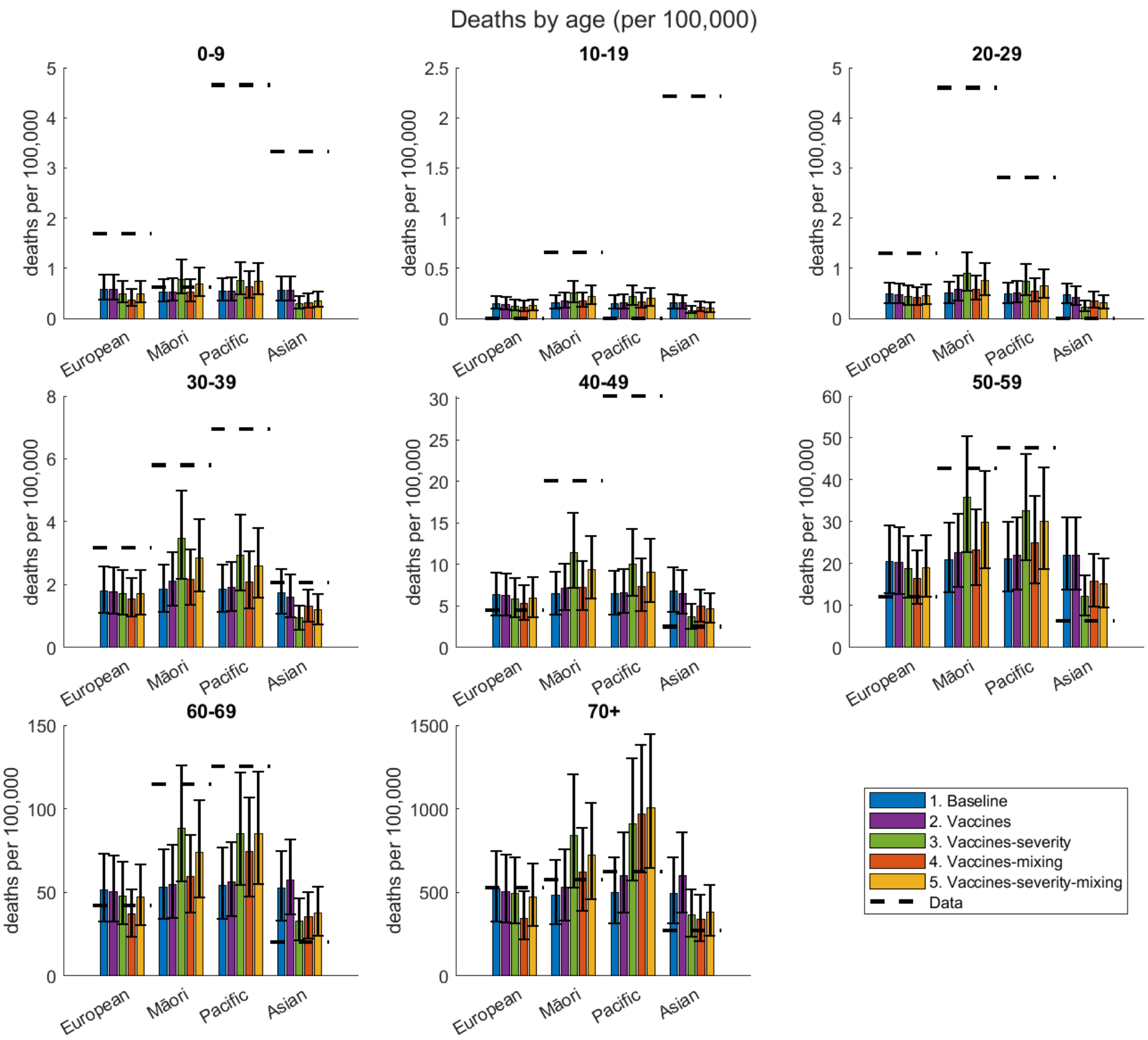

**Supplementary Figure 7.** Cumulative number of Covid-19 deaths per 100,000 people in 10-year age groups from 1 January 2022 to 30 June 2023, comparing model scenarios (coloured bars) and data (horizontal red dashed lines). Coloured bars show the median; error bars show the 95% CrI. Note that data on death rates in younger age groups are based on very small numbers of deaths and hence the effect of stochastic fluctuations is relatively large.

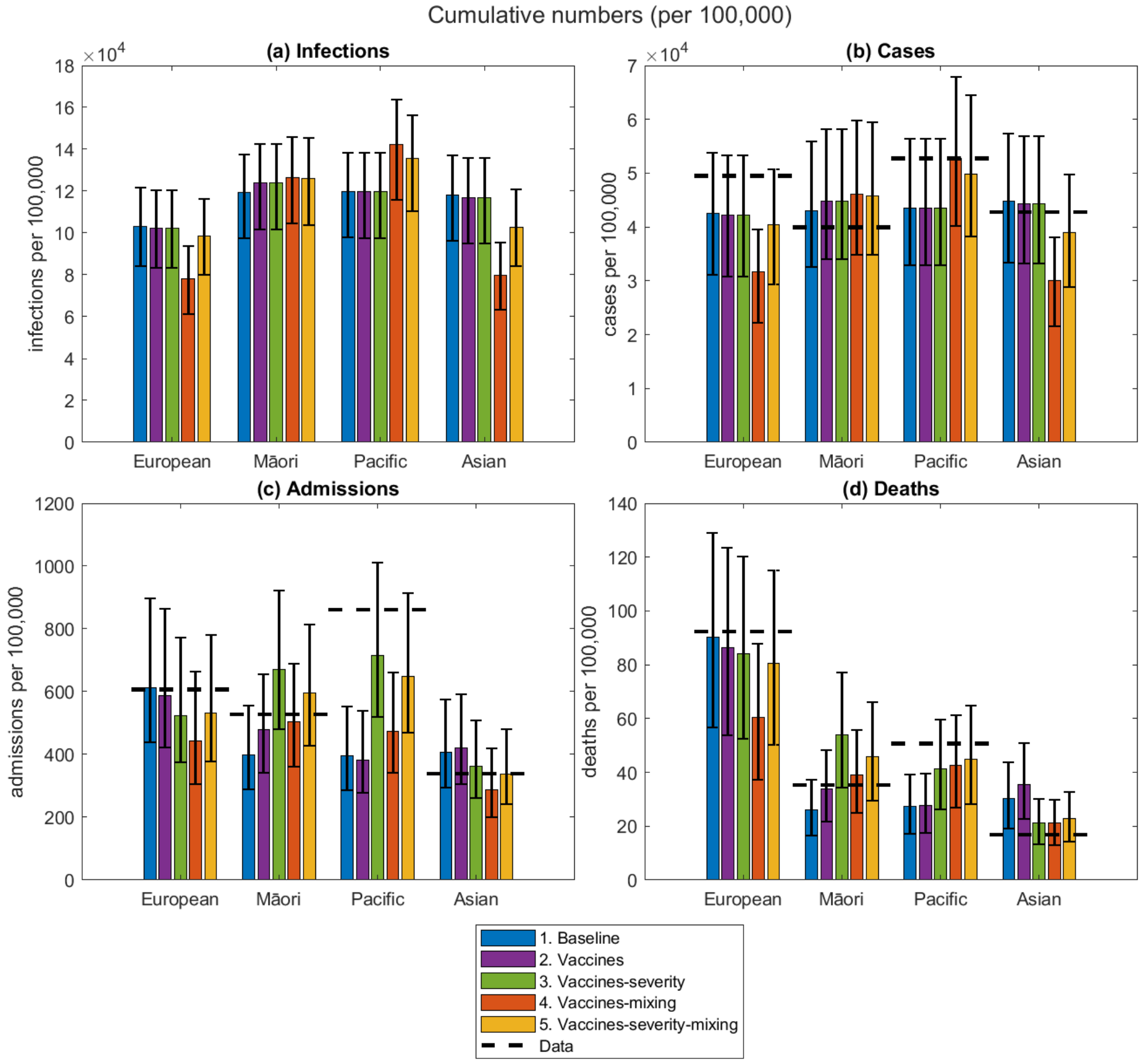

**Supplementary Figure 8.** Results of using the Statistics NZ population projections instead of the Health Service User (HSU) dataset to specify the population by ethnicity and age group. Cumulative number of SARS-CoV-2 infections, reported Covid-19 cases, Covid-19 hospital admissions and Covid-19 deaths per 100,000 people for each ethnicity, aggregated over all ages, from 1 January 2022 to 30 June 2023, comparing model scenarios (coloured bars) and data (horizontal red dashed lines). Coloured bars show the median; error bars show the 95% CrI.